\documentclass[
 preprint,
 superscriptaddress,
 amsmath,amssymb,
 aps,
 prl,
 longbibliography,
 showpacs
]{revtex4-1}
\usepackage{graphicx}
\usepackage{braket}
\usepackage{dcolumn}
\usepackage{bm}
\usepackage{epstopdf}

\begin{document}

\title{Electron and hole g tensors of neutral and charged excitons in single quantum dots by high-resolution photocurrent spectroscopy}
\author{Shiyao Wu}
\thanks{Contributed equally to this work.}
\author{Kai Peng}
\thanks{Contributed equally to this work.}
\affiliation{Beijing National Laboratory for Condensed Matter Physics, Institute of Physics, Chinese Academy of Sciences, Beijing 100190, China}
\affiliation{CAS Center for Excellence in Topological Quantum Computation and School of Physical Sciences, University of Chinese Academy of Sciences, Beijing 100049, China}
\author{Xin Xie}
\author{Jingnan Yang}
\author{Shan Xiao}
\author{Feilong Song}
\author{Jianchen Dang}
\author{Sibai Sun}
\author{Longlong Yang}
\affiliation{Beijing National Laboratory for Condensed Matter Physics, Institute of Physics, Chinese Academy of Sciences, Beijing 100190, China}
\affiliation{CAS Center for Excellence in Topological Quantum Computation and School of Physical Sciences, University of Chinese Academy of Sciences, Beijing 100049, China}

\author{Yunuan Wang}
\affiliation{Beijing National Laboratory for Condensed Matter Physics, Institute of Physics, Chinese Academy of Sciences, Beijing 100190, China}
\affiliation{Key Laboratory of Luminescence and Optical Information, Ministry of Education, Beijing Jiaotong University, Beijing 100044, China}
\author{Shushu Shi}
\author{Jiongji He}
\author{Zhanchun Zuo}
\affiliation{Beijing National Laboratory for Condensed Matter Physics, Institute of Physics, Chinese Academy of Sciences, Beijing 100190, China}
\affiliation{CAS Center for Excellence in Topological Quantum Computation and School of Physical Sciences, University of Chinese Academy of Sciences, Beijing 100049, China}

\author{Xiulai Xu}
\email{xlxu@iphy.ac.cn}
\affiliation{Beijing National Laboratory for Condensed Matter Physics, Institute of Physics, Chinese Academy of Sciences, Beijing 100190, China}
\affiliation{CAS Center for Excellence in Topological Quantum Computation and School of Physical Sciences, University of Chinese Academy of Sciences, Beijing 100049, China}
\affiliation{Songshan Lake Materials Laboratory, Dongguan, Guangdong 523808, China}

\date{\today}




\date{\today}

\begin{abstract}

We report a high-resolution photocurrent (PC) spectroscopy of a single self-assembled InAs/GaAs quantum dot (QD) embedded in an n-i-Schottky device with an applied vector magnetic field. The PC spectra of positively charged exciton (X$^+$) and neutral exciton (X$^0$) are obtained by two-color resonant excitation. With an applied magnetic field in Voigt geometry, the double $\Lambda$ energy level structure of X$^+$ and the dark states of X$^0$ are observed in PC spectra clearly. In Faraday geometry, the PC amplitude of X$^+$ decreases and then quenches with the increasing of the magnetic field, which provides a new way to determine the relative sign of the electron and the hole g-factors. With an applied vector magnetic field, the electron and the hole g-factor tensors of X$^+$ and X$^0$ are obtained. The anisotropy of the hole g-factors of both X$^+$ and X$^0$ is larger than that of the electron.


\end{abstract}

\maketitle

\section{\label{sec:level1}Introduction}

Electron and hole spins in semiconductor material have long been considered a building block in quantum information processing and quantum computation \cite{DiVincenzo2002}. Three-dimensionally confined quantum dots (QDs) provide a long coherent time for spin states allowing the single spin state manipulation \cite{PhysRevA.57.120,PhysRevB.69.075320,Press2008}. During the coherent control of the spins in QDs, a magnetic field is usually used to modulate the properties of spins and energy levels \cite{Gaudreau2012,Awschalom1174}. The g-factor, a very important parameter describing the coupling of spins to a fixed magnetic field on the QDs \cite{Bennett2013,Kato1201}, has been intensively studied in various systems. Generally, the g-factor shows anisotropy in different directions, and can be described as a g-factor tensor. Some efforts have been made to characterize the exciton g-factor \cite{apl-1.2937305, PhysRevB.72.201307,PhysRevB.83.041307} and the g-factor tensor of confined carriers in QDs through optical methods \cite{PhysRevLett.107.166604, PhysRevLett.97.236402,apl2011,Wu2019,PhysRevB.93.125302}.
In self-assemble QDs, the electron and hole g-factors are related to the size, shape and composition \cite{PhysRevB.83.161303, PhysRevB.70.235337}, and can be tuned by the external field\cite{PhysRevB.94.245301,PhysRevB.99.195305,PhysRevB.85.155310,aplKlotz,PhysRevB.71.205301,Weidong-apl}. Additionally, the g-factors usually vary for different excitons due to the Coulomb interactions \cite{Bennett2013}. However, there are few systematic works showing the electron and hole g-factor tensors of different excitons \cite{PhysRevB.84.195305, PhysRevB.72.161312}, which limits the full description of the electronic properties of different excitons in single QDs.

Photocurrent (PC) spectroscopy of a single QD embedded in a diode has been proved an effective way to read out the charge \cite{Zrenner2002} and the spin \cite{PhysRevLett.100.197401,PhysRevB.90.241303} information from a single QD directly with high detection efficiency\cite{Ramsay2010}. Compared with photoluminescence (PL) by non-resonant pumping, the PC spectroscopy has a higher resolution with continuous-wave (CW) resonant pumping, and can be used in the coherent control with pulsed pumping \cite{PhysRevB.76.041301,PhysRevLett.114.137401,PhysRevLett.106.067401,doi:10.1063/1.5020364,PhysRevLett.96.037402,PhysRevLett.102.207401}. On the other hand, the detection of PC signal doesn't need to filter the pumping laser like resonant fluorescence, which is convenient to research the spin properties. While for the resonant fluorescence, a cross-polarization configuration is usually constructed to eliminate the resonant laser background \cite{vamivakas2009spin,he2013demand}, which limits the extraction of polarized signal for spin information.

In this work, we report on g-factor tensors of positively charged exciton X$^+$ and neutral exciton X$^0$ in a single self-assemble InAs/GaAs QD through a high-resolution PC spectroscopy under the vector magnetic fields. The QDs are embedded in the intrinsic region of an n-i-Schottky diode with a two-dimensional electron gas (2DEG) beneath. The two-color CW narrow-bandwidth ($\sim$1MHz) lasers are used to perform the high-resolution PC measurement of X$^+$. With an applied vector magnetic field, X$^+$ and X$^0$ show Zeeman splitting and dark exciton states. The electron and hole g-factor tensors of X$^+$ and X$^0$ are obtained through the PC spectra. Besides, the PC signal of X$^+$ quenching at a large magnetic field in Faraday geometry reveals the opposite signs of electron and hole g-factors. The hole g-factors show larger anisotropy than those of electrons for both X$^+$ and X$^0$, revealing the hole g tensor is more sensitive to the geometry of the QDs. This work shows PC spectroscopy is an effective and convenient way to characterize the g-factors with high resolution.

\begin{figure}
\includegraphics[scale=0.5]{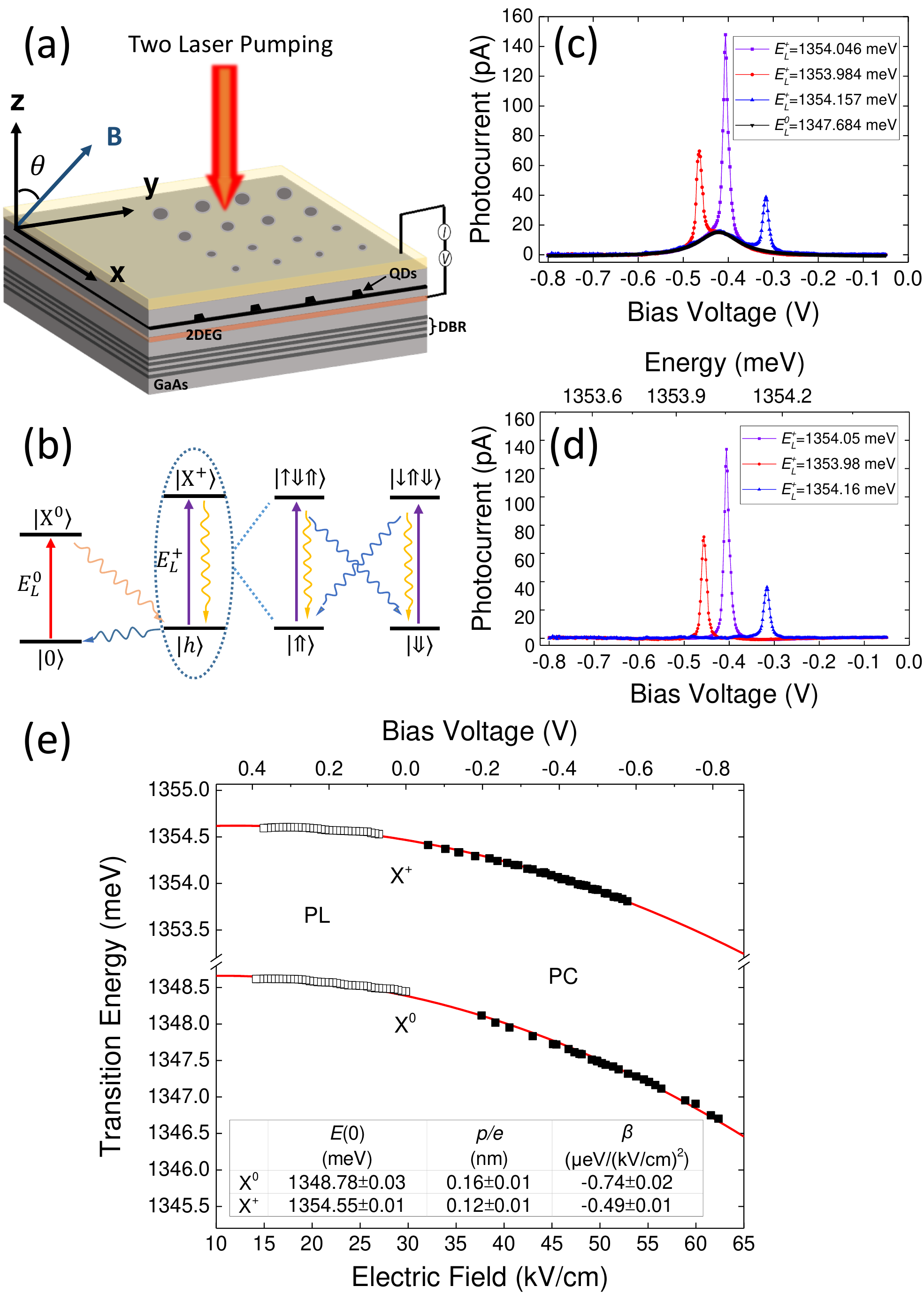}
\caption{\label{fig:1}(a) The schematic diagram of the n-i-Schottky device and the PC measurement under a vector magnetic field. Here $\theta$ is the angle between the magnetic field (B) and the growth axis of the QDs (z direction). (b) The two-color excitation scheme used for X$^+$ PC measurements. Solid and wave arrows represent the excitations and the tunnelling processes of carriers, respectively. (c) PC spectra of X$^0$ and X$^+$ under the two-color excitation. The black solid points represent the pure X$^0$ PC spectrum with \emph{E$^0_L$}=1347.684 meV; The red, purple and blue solid points represent the measured PC spectra of X$^+$ for different energies of the second laser \emph{E$^+_L$}. (d) The subtracted PC spectra of X$^+$ for different energies of the second laser. The top x-axis represents the Stark shift of X$^+$. (e) Stark shifts of X$^0$ and X$^+$ as a function of the electric field and the bias voltage obtained from the single QD PL (empty square) and PC (solid square) spectra. The solid red lines represent quadratic fittings by quantum confined Stark effect for X$^0$ and X$^+$. Inset: The fitting parameters for \emph{E(0)}, \emph{p}, and $\beta$ are given in the table.}
\end{figure}

\section{\label{sec:level1}EXPERIMENTAL DETAILS}

The n-i-Schottky device was designed and fabricated for performing PC measurement of single QDs. The schematic diagram of the device is shown in Fig.~\ref{fig:1}(a). The Schottky contact was formed by evaporating a 10-nm semitransparent titanium at the surface, followed by an Al mask with apertures of about 1-3 $\mu$m to isolates single QDs. A 2DEG formed by Si $\delta$-doped GaAs layer (Nd = $5 \times 10^{12} cm^{-2}$),  as the Fermi sea of electron in the device, is connected with Cr/Au bond pads on top of an alloyed (Au, Ge)Ni ohmic contact layer \cite{apl1999Toshiba, prl2004Bernd}. To enhance the photon collection efficiency, a distributed Bragg reflector (DBR) of 13-pairs of Al$_{0.94}$Ga$_{0.06}$As/GaAs (67/71 nm) was grown at the bottom of the structure \cite{Nature2019Richard, santori2002indistinguishable}. The vertical electric field can be applied on the QDs as \emph{F}=(\emph{V}$_i$-\emph{V}$_b$)/\emph{d}, where \emph{V}$_i$ is the built-in potential (0.74 V for this device), \emph{V}$_b$ is the applied bias voltage and \emph{d} is the distance between the Schottky contact and the 2DEG. Further details on the design and the fabrication of the device can be found in Ref \onlinecite{PhysRevB.83.075306}.

To perform resonant excitation of QDs, two tunable narrow-bandwidth ($\sim$1 MHz) external cavity diode lasers in Littrow configuration were furnished. The non-resonant PL measurement was excited by a 650-nm semiconductor laser. A confocal microscope objective with a large numerical aperture of NA=0.82 was used to perform micro-PL and PC measurement for single QDs. The PL singals were collected and dispersed through a 0.55-m spectrometer with a spectral resolution of about 60 $\mu$eV. To perform polarization-resolved PL and PC measurements, half-wave plates and polarizers were added in the pumping and collection paths. For all the measurements in this work, the device was maintained at 4.2 K on an xyz piezoelectric stage in a helium gas exchange cryostat equipped with superconducting magnets to supply a vector magnetic field up to 4 T. A semiconductor analyzer with a high current resolution (10 fA) was used for the electrical measurements.

\section{\label{sec:level1}Results and disscussion}


The PC spectra are obtained through resonant excitations of X$^0$ and X$^+$. Firstly, the X$^0$ is excited by a laser \emph{E$_L^0$} resonantly. The electron will tunnel out of the QDs preferentially due to the small effective mass, leaving a single heavy hole in the valance band, as shown in Fig.~\ref{fig:1}(b). Through pumping-power-dependent PC measurements, the analysis of linewidth gives the typical electron tunneling time as several picoseconds with tunnel barrier height of about 60 meV \cite{2019apl,doi:10.1063/1.3633216}. The hole will then tunnel out of the QDs and contribute to a X$^0$ PC signal, as shown by the black solid points in Fig.~\ref{fig:1}(c). In this QD, the hole tunneling time is measured as 3.96 ns with tunnel barrier of 45.51 meV by the analysis of PC amplitude in the power-dependent measurement \cite{PhysRevApplied.11.024015}. Meanwhile, if a second laser \emph{E$_L^+$} resonant with X$^+$ is applied on the QD, the hole state $\ket{h}$, as a initial state, can be excited to X$^+$. After the electron tunnels out of the QDs, one hole will tunnel out first due to the enhanced tunneling rate induced by the Coulomb repulsion between two holes. Here, the Coulomb interaction lowers the tunnel barrier to 37.46 meV, and accelerate the hole tunneling about 30 times faster as 0.14 ns \cite{PhysRevApplied.11.024015}. The remaining hole can either tunnel out of the QDs or be excited to X$^+$ again. Here, the linear-polarized lasers are used to excite both X$^0$ and X$^+$. So all the spin states of X$^+$ and X$^0$ can be excited equally. No matter in which hole spin state, the system can be repumped to X$^+$ again, as shown in the right panel of Fig.~\ref{fig:1}(b). The Coulomb-induced hole tunneling acceleration and the reuse of the hole state contribute to a much larger PC signal of X$^+$ than X$^0$ as shown in Fig.~\ref{fig:1}(c). A detailed discussion of this mechanism can be found in our previous work \cite{PhysRevApplied.11.024015}. By tuning the excitation energy of \emph{E$_L^+$}, the X$^+$ PC peaks appear at different bias voltages accordingly, as shown by the colored curves in Fig.~\ref{fig:1}(c). Through a simple subtraction at relatively low pumping power condition, the pure X$^+$ PC spectra are obtained, as shown in Fig.~\ref{fig:1}(d). Compared with X$^0$, the PC spectrum of X$^+$ has a narrow linewidth due to the Coulomb attraction prolonging the electron tunneling time. The linewidth of the X$^+$ peaks in PC spectra in Fig.~\ref{fig:1}(d) are all about 15 $\mu$eV, which offers a higher resolution than X$^0$ with a linewidth of 230 $\mu$eV.

The exciton transition energy varies with the electric field (bias voltage) is described by the Stark shift with the expression \emph{E(F)}=\emph{E}(0)+\emph{pF}+$\beta$\emph{F}$^2$, where \emph{E}(0) is the transition energy without the external electric field, \emph{p} is the permanent dipole moment and $\beta$ is the polarizability of electron-hole wavefunctions. By tuning the pumping laser wavelength, a series of PC spectra are obtained for X$^0$ and X$^+$. Combining the the series of PL and PC spectra, the Stark shifts are shown in Fig.~\ref{fig:1}(e). With fitting results in the inset of Fig.~\ref{fig:1}(e), the functions between the electric field (bias voltage) applied on the QD and the transition energies of X$^0$ and X$^+$ can be determined precisely. According to the determined function, the corresponding energy of X$^+$ can be derived, as shown in the top x-axis in Fig.~\ref{fig:1}(d). Furthermore, the fitting parameters in Fig.~\ref{fig:1}(e) can also provide the dipole moment and polarizability of X$^0$ and X$^+$ precisely \cite{PhysRevB.95.201304}.


\begin{figure}
\includegraphics[scale=0.5]{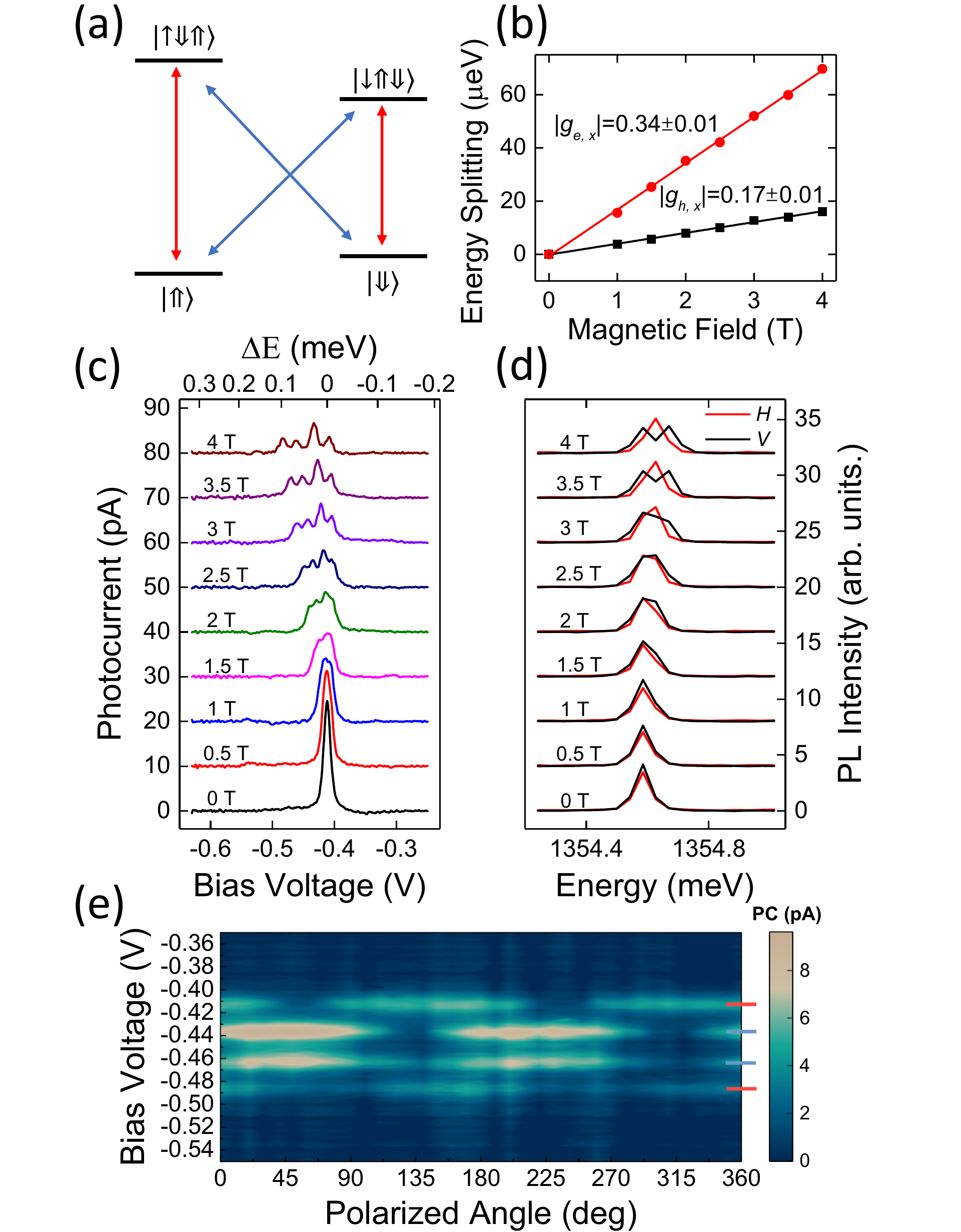}
\caption{\label{fig:2}(a) The energy level structure of X$^+$ under the magnetic field. Here the two inner transitions (blue solid arrows) with the angular momentum of $M=\pm2$ are dark state transitions. The two outer transitions (red solid arrows) with $M=\pm1$ are bright state transitions. (b) Zeeman splitting of $\ket{X^{+}}$ and $\ket{h}$ states in Voigt geometry. (c) PC spectra of X$^+$ with a magnetic field from 0 to 4 T in Voigt geometry. The spectra are shifted for clarity. (d) The polarization-resolved PL spectra of X$^+$ with the magnetic field from 0 to 4 T in Voigt geometry. The spectra are shifted for clarity. (e) The polarization-resolved PC spectra of X$^+$ under 4-T Voigt magnetic field, where the blue and red short lines at right side are eyes guide for the dark states and bright states respectively.}
\end{figure}


\begin{figure*}
\includegraphics[scale=0.4]{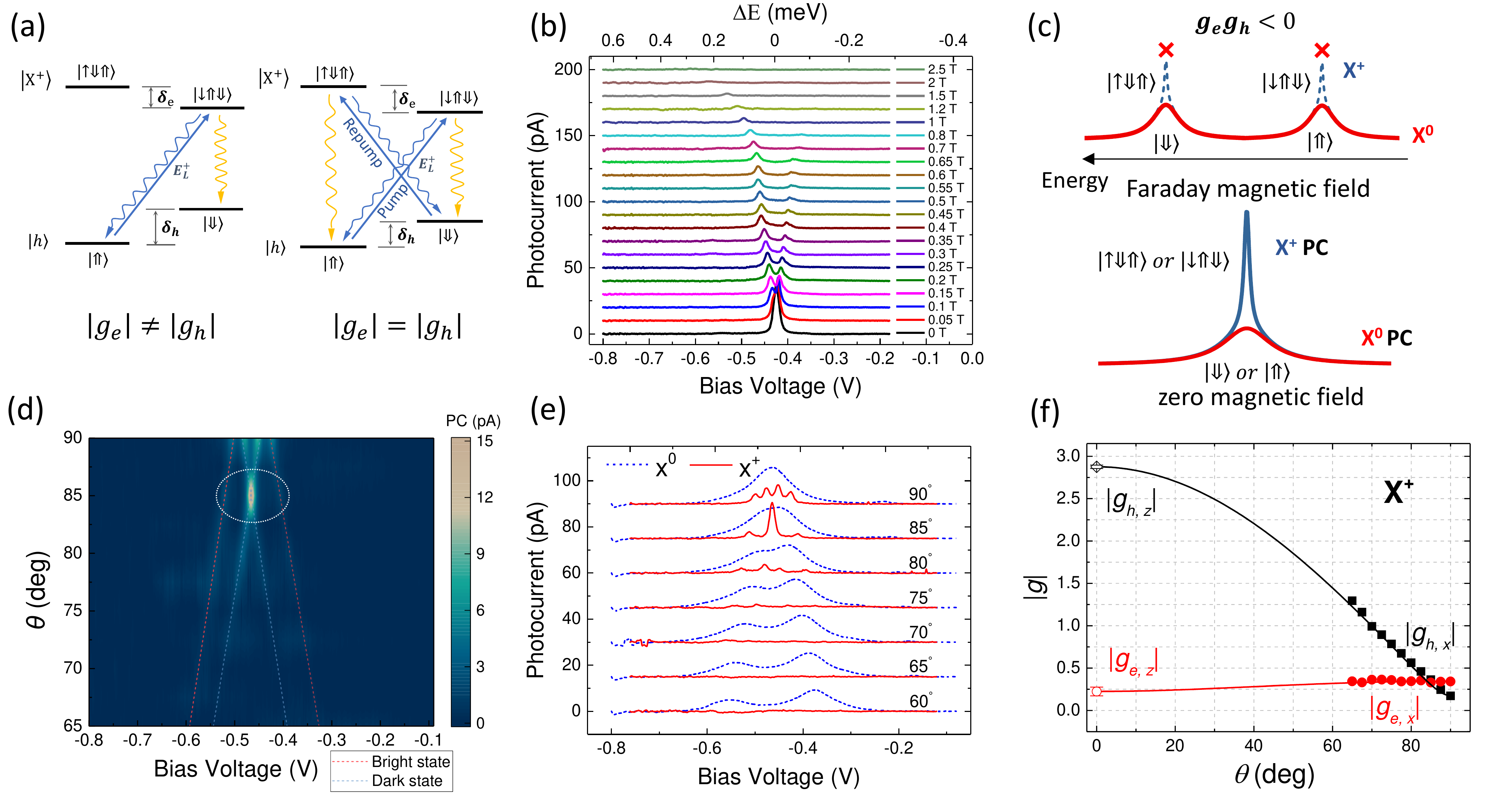}
\caption{\label{fig:3} (a) The schematic diagram of the carrier transitions and decays in X$^+$ under a magnetic field with $|g_e|\neq|g_h|$ in the left panel and $|g_e|=|g_h|$ in the right panel. The solid and wavy lines represent the excitation and the tunneling process, respectively. (b) PC spectra of  X$^+$ with the magnetic field in Faraday geometry from 0 to 2.5 T. The top x-axis represents the energy shift compared with the X$^+$ transition energy at zero magnetic field. The spectra are shifted for clarity. (c) The schematic diagram of X$^+$ and X$^0$ PC spectra under zero (bottom panel) and Faraday (top panel) magnetic field with the condition of $g_{e}g_{h}<0$. Here the red and blue curves represent the PC spectra of X$^0$ and X$^+$, respectively. (d) Magneto-PC mapping of X$^+$ under a 4-T vector magnetic field with angle from $90^{\circ}$ to $65^{\circ}$. Here the red and blue dashed lines are the eye guide for the bright and dark state transitions of X$^+$, respectively. (e) The PC spectra of  X$^+$ and  X$^0$ under a 4-T vector magnetic field with $\theta$ from $90^{\circ}$ to $60^{\circ}$. The red solid curves are the PC spectra of  X$^+$ and the blue dashed curves are the PC spectra of  X$^0$. (f) Electron and hole g-factors of X$^+$ as a function of the angle $\theta$. The solid dots are the extracted electron and hole g-factors with error bars of about 0.02. The red and black dot at $\theta=0^{\circ}$ show the derived $\vert{g_{e,z}}\vert$ and $\vert{g_{h,z}}\vert$ by fitting the extract data through Eq.~\eqref{1} (solid curves) with the condition of $|{g_{e,z}}|+|{g_{h,z}}|=3.10\pm0.02$.}
\end{figure*}

In order to obtain the electron and hole g-factor tensor, the PC spectrum under the magnetic field in Voigt geometry is first measured. When a magnetic field is applied perpendicular to the sample growth direction (z direction), namely, in Voigt geometry, the spin degeneracy of electron and hole will be lifted, which induces the Zeeman splitting of $\ket{X^{+}}$ and $\ket{h}$ states and forms a double $\Lambda$ structure, as presented in Fig.~\ref{fig:2}(a). Generally, due to the strain-induced large splitting between heavy- and light-hole states, only heavy hole ($\ket{\Uparrow}$ or $\ket{\Downarrow}$) with the angular momentum $J_{h,z}=\pm3/2$ is considered in self-assemble QDs \cite{PhysRevB.41.5283, PhysRevB.65.195315}. For the electron ($\ket{\uparrow}$ or $\ket{\downarrow}$) in the bottom of the conduction band, the total angular momentum is equal to the spin momentum $S_{e,z}=\pm1/2$. Therefore, a total angular momentum of the electron-hole pairs can be defined as $M=J_{h,z}+S_{e,z}$. Transitions with $M=\pm1$ satisfy the selection rule, are defined as bright states; while transitions with $M=\pm2$ cannot couple to the light field, are defined as dark states. For X$^+$, the value of M can be defined as the change between the initial and final states. Here, the transitions $\ket{\uparrow\Downarrow\Uparrow}\leftrightarrow\ket{\Uparrow}$ and $\ket{\downarrow\Uparrow\Downarrow}\leftrightarrow\ket{\Downarrow}$ (red arrows) have the angular momentum of $M=\pm1$, which satisfy the selection rules, are defined as bright states.  While the transitions $\ket{\uparrow\Downarrow\Uparrow}\leftrightarrow\ket{\Downarrow}$ and $\ket{\downarrow\Uparrow\Downarrow}\leftrightarrow\ket{\Uparrow}$ (blue arrows) with the angular momentum of $M=\pm2$, usually defined as dark states, are optically forbidden. However, the magnetic field in Voigt geometry will destruct the rotation symmetry and couple the bright and dark states, resulting in the optical access and linear polarization of the dark states.

Fig.~\ref{fig:2}(c) shows the corresponding PC spectra of X$^+$ with the appearance of the double $\Lambda$ structure, in which the top x-axis represents the energy shift towards 0 T. Here, the excitation laser is set linear-polarized with the polarization direction along $45^\circ$ respect to the applied Voigt magnetic field, so as to excite the four transitions of X$^+$ equally. It is worth noting that because of the fixed energy of the pumping laser, the energy level of X$^+$ is tuned lower to match the laser energy with an additional energy shift induced by the diamagnetic effect and the Zeeman splitting. As a result, the peaks of the PC spectra move to the negative bias voltage (high electric field) direction at a high magnetic field \cite{PhysRevApplied.8.064018}. Here, due to the giant PC enhancement of X$^+$ caused by the reuse of hole state is destroyed by the magnetic field, the PC amplitudes in Fig.~\ref{fig:2}(c) decrease with the increase of the magnetic field. In order to illustrate this process, we first assume the laser excite the transition $\ket{\Uparrow}\rightarrow\ket{\downarrow\Uparrow\Downarrow}$, as shown in the left panel in Fig.~\ref{fig:3}(a). After one pair of electron and hole in X$^+$ tunnel out, the system will decay to $\ket{\Uparrow}$ or $\ket{\Downarrow}$. As it decays to  $\ket{\Uparrow}$, the laser can reuse the hole state and excite the system back to $\ket{\downarrow\Uparrow\Downarrow}$. While once it decays to $\ket{\Downarrow}$, due to the detuning between laser energy and the transition energies from $\ket{\Downarrow}$ to X$^+$, the system will stay in this state until the hole tunnels out. As a result, the enhancement mechanism is destroyed because of the Zeeman splitting, which causes the decrease of the X$^+$ PC amplitude under the magnetic field. In contrast to the PC spectra, we cannot distinguish any splitting from the non-resonant excitation PL spectra but only a linewidth broadening. Even through the polarization-resolved PL measurements, only three peaks can be observed at 4 T, as shown in Fig.~\ref{fig:2}(d). Clearly, the PC spectroscopy could offer a better resolution than the PL in our experiment.

The polarization-resolved PC spectra of X$^+$ in a 4-T Voigt configuration magnetic field are shown in Fig.~\ref{fig:2}(e), which reveals the orthogonal linear polarizations of the four transitions of X$^+$ clearly and proves the formation of the double $\Lambda$ energy level structure again. The two outer PC peaks with polarizations orthogonal to the Voigt magnetic field, as marked by red short lines, are the transitions labeled as red arrows in Fig.~\ref{fig:2}(a). While the two inner PC peaks, as guided by blue short lines, are the corresponding emissions of the two diagonal transitions labeled as blue arrows in Fig.~\ref{fig:2}(a), which have the polarization perpendicular to the two outer transitions. Through the Lorentzian fitting of the PC peaks in Fig.~\ref{fig:2}(c), the splittings of the $\ket{X^{+}}$ and $\ket{h}$ as a function of the magnetic field in Voigt geometry are extracted, as the red and black dots shown in Fig.~\ref{fig:2}(b). For this QD, the in-plane g-factors of electron and hole are $\vert{g_{e,x}}\vert=0.34\pm0.01$ and $\vert{g_{h,x}}\vert=0.17\pm0.01$, respectively.


\begin{figure}
\includegraphics[scale=0.5]{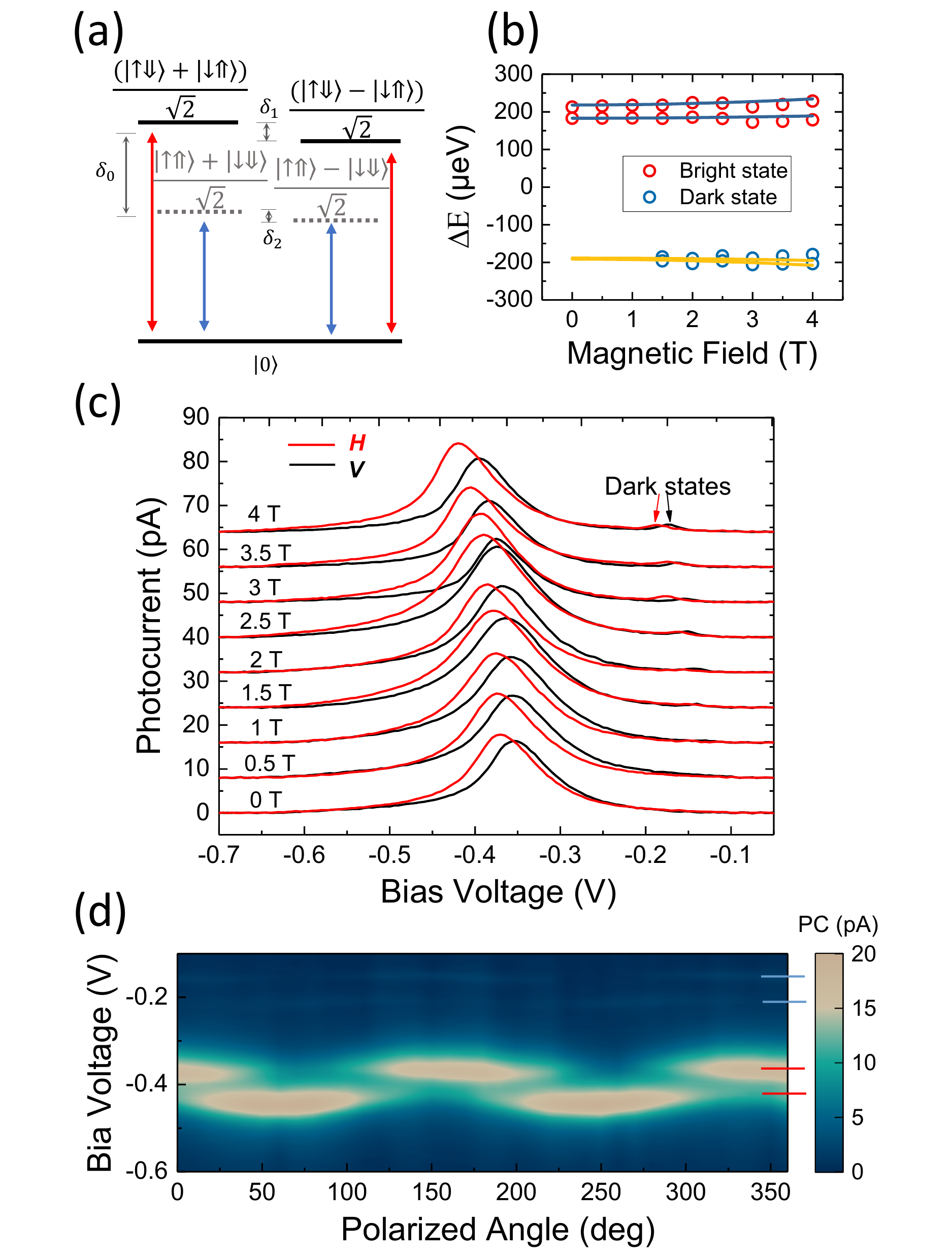}
\caption{\label{fig:4} (a) The energy level structure of X$^0$ with the magnetic field in Voigt geometry. Here $\delta_{0}$, $\delta_{1}$ and $\delta_{2}$ represent the energy differences between the bright states and the dark states and the FSS of bright and dark states respectively, which can be referred as the exchange parameters. (b) Zeeman splitting of bright and dark states with a magnetic field in Voigt geometry. The diamagnetic shift is deduced. Here the circles are the extracted data, and the solid curves are the fitted results with considering the exchange parameter ($\delta_0$, $\delta_1$ and $\delta_2$). (c) The polarization-resolved PC spectra of X$^0$ with a magnetic field from 0 to 4 T in Voigt geometry. The spectra are shifted for clarity. (d) The linear-polarization-resolved PC spectra at 4 T in Voigt geometry. The blue and red short lines are the eyes guide for the dark and bright states of X$^0$. }
\end{figure}

To determine the z component of the electron and hole g-factor tensor, the PC spectra measurement are performed with the magnetic field in both Faraday geometry and intermediate geometry from Faraday to Voigt geometry, as shown in Fig.~\ref{fig:3}. In the Faraday geometry, i.e. the magnetic field along z direction, the PC spectra of X$^+$ split into two peaks with opposite circular polarizations, as shown in Fig.~\ref{fig:3}(b). The observed Zeeman splitting is determined by the electron and hole g-factor, which can be expressed as $\triangle{E _{Zeeman}}=\mu_{B}(|{g_{e,z}}-{g_{h,z}}|)B$. By extracting the Zeeman splitting from Fig.~\ref{fig:3}(b), the exciton g-factor $|{g_{e,z}}-{g_{h,z}}|$ is deduced as $3.10\pm0.02$. Due to the same reason with what happens in Voigt geometry, the PC amplitude gets weaken with the increase of the magnetic field. When the magnetic field is above 1.5 T, the PC amplitude starts quenching which implies the opposite signs of electron and hole g-factor. To explain this phenomenon we need to consider the zero magnetic field first. At zero magnetic field, the PC spectrum of X$^0$ offers a hole background after electron's tunneling. The red solid curve in the bottom panel of Fig.~\ref{fig:3}(c) represents the hole background with the same probability of spin-up and spin-down. With the second laser \emph{E$_L^+$} on, the two-color excitation will excite the system to X$^+$, as the blue solid curve shown in the bottom panel of Fig.~\ref{fig:3}(c). At this time, due to the superposition of different spin hole backgrounds, there is no need to consider spin selection. While in a magnetic field with Faraday geometry, due to the large g-factor of hole, the hole background offered by X$^0$ splits into two peaks when the magnetic filed is above 1.5 T, as shown by the red solid curve shown in the top panel in Fig.~\ref{fig:3}(c). Here, for a better interpretation, we assume that the hole background on the high (low) energy side provides the holes with spin down $\ket{\Downarrow}$ (up $\ket{\Uparrow}$). If the sign of the electron g-factor is opposite, then X$^+$ will have the state $\ket{\uparrow\Downarrow\Uparrow}$ ($\ket{\downarrow\Uparrow\Downarrow}$) at the corresponding energy, as shown by the blue dashed curves shown in the top panel in Fig.~\ref{fig:3}(c). However, due to the the selection rule, this transition $\ket{\Downarrow}\longleftrightarrow\ket{\uparrow\Downarrow\Uparrow}$ ($\ket{\Uparrow}\longleftrightarrow\ket{\downarrow\Uparrow\Downarrow}$) with angular momentum $M=\pm2$ will be forbidden under Faraday magnetic field, leading to the PC amplitude of X$^+$ quenches with a magnetic field up to 1.5 T. Therefore, the two-color excitation scheme of X$^+$ provides a way to get the relative signs of the electron and hole g-factor, and the g-factor of X$^+$ can be represented as $|{g_{e,z}}|+|{g_{h,z}}|$ through the opposite sign of ${g_{e,z}}$ and ${g_{h,z}}$. It is worth noting that due to the larger electric field of the high energy branch, the PC amplitude is larger than that at lower energy branch in Fig.~\ref{fig:3}(b). As a result, the PC signals quench faster at lower energy branch.

The PC spectra under a vector magnetic field were performed as a function of $\theta$, as defined in Fig.~\ref{fig:1}(a). The color-filled PC spectra of X$^+$ under a 4-T vector magnetic field with $\theta$ range from $90^{\circ}$ to $65^{\circ}$ is shown in Fig.~\ref{fig:3}(d). As the angle $\theta$ decreases, i.e. the magnetic field from Voigt to Faraday geometry, the splitting of the outer transitions increases and the two inner transitions lines cross each other. An interesting feature is the PC enhancement of the inner transitions at $\theta=85^{\circ}$. This enhancement can be ascribed to the repumping of the hole state\cite{PhysRevB.90.121402}, as the schematic diagram shown in the right panel in Fig.~\ref{fig:3}(a).  Different from the case for $|g_e|\neq|g_h|$ (left panel in Fig.~\ref{fig:3}(a)), the absolute value of electron g-factor is equal to that of the hole at $\theta=85^{\circ}$, resulting in the same energy of diagonal transitions. Therefore, the laser resonant with both diagonal transitions can repump the hole state to X$^+$ state, no matter which hole spin state ($\ket{\Downarrow}$ or $\ket{\Uparrow}$) it is in. Thus the reuse of hole state can be restored, and the PC amplitude gets much larger, as shown in Fig.~\ref{fig:3}(d) and (e) clearly. The PC amplitude gets weaker with $\theta$ decreases, which is due to the same reason as that with large Faraday magnetic fields discussed before. As shown in Fig.~\ref{fig:3}(e), for $\theta$ from $90^{\circ}$ to $60^{\circ}$, the PC spectrum of X$^0$ splits into two peaks (blue dashed lines). The forbidden transitions caused by the opposite signs of the electron and hole g-factors weaken the X$^+$ PC (red solid lines) amplitudes and eventually make it disappear when the PC spectrum of X$^0$ is totally separated.

The behavior of X$^+$ under a vector magnetic field can be described by g-factors of electron and hole as a function of $\theta$:
\begin{equation}
g_{e,h}(\theta)=\sqrt{(g_{e,h,x}cos\theta)^2+(g_{e,h,z}sin\theta)^2}\ .
\label{1}
\end{equation}
According to Eq.~(\ref{1}), the extracted g-factors of electron and hole can be fitted, as shown in Fig.~\ref{fig:3}(f). Here, the g-factor of X$^+$ ($|g_e,z|+|g_h,z|=3.10\pm0.02$) is also used as a parameter to fit the g-factor tensor in Fig.~\ref{fig:3}(f). The hole g-factor shows a larger anisotropy than electron, consistent with other works \cite{PhysRevB.92.165307,PhysRevB.93.125302}. The anisotropy factors of electron and hole g-factors can be described as $P_{e, x-z}=({g_{e, x}-g_{e, z}})/({g_{e, x}+g_{e, z}})=(17.91\pm0.01) \%$ and $P_{h, x-z}= ({g_{h, x}-g_{h, z}})/({g_{h, x}+g_{h, z}})=(88.55\pm0.02) \%$, as shown in Tables~\ref{tbl:example}. This can be ascribed to the anisotropic QD shape leading to the different spatial real-space wave function distributions and the non-zero orbital momenta carried by hole states \cite{PhysRevB.93.035311,PhysRevB.75.033316,Sheng2008}. Previous works with coherent population trapping by resonant fluorescence spectroscopy can measure the g-factors of X$^+$ with very high resolution \cite{PhysRevLett.112.107401,Prechtel2016}, while it also needs sophisticated device fabrication and experimental technique. Here, the detection of the PC spectroscopy provides a convenient way to investigate the spin properties.




Different from X$^+$, due to electron-hole exchange interaction and the asymmetry of the QD structure, the energy degeneracy of the neutral exciton X$^0$ states without magnetic field is lifted by the fine structure splitting (FSS). A systematic description can be found in Ref.\onlinecite{PhysRevB.65.195315}. For the X$^0$ consisting of an electron and a heavy hole, the exction states $\ket{\uparrow\Downarrow}$ and $\ket{\downarrow\Uparrow}$ are mixed as the new basis with total angular momentum $M=\pm1$: $\dfrac{\ket{\uparrow\Downarrow}+\ket{\uparrow\Downarrow}}{\sqrt{2}}$ and $\dfrac{\ket{\uparrow\Downarrow}-\ket{\uparrow\Downarrow}}{\sqrt{2}}$, which satisfy the selection rule and are defined as the bright states; while the exciton states $\ket{\uparrow\Uparrow}$ and $\ket{\downarrow\Downarrow}$ are mixed as: $\dfrac{\ket{\uparrow\Uparrow}+\ket{\downarrow\Downarrow}}{\sqrt{2}}$ and $\dfrac{\ket{\uparrow\Uparrow}-\ket{\Downarrow\Downarrow}}{\sqrt{2}}$ with total angular momentum $M=\pm2$, forbidden by the selection rule, are defined as the dark states, as shown in Fig.~\ref{fig:4}(a). Additionally, the mixture of the exciton states induces the orthogonally linear polarizations, which enable us to distinguish the FSS by polarization-resolved PC spectra. The FSS of the bright states $\delta_{1}$ is usually larger than that of the dark states $\delta_{2}$. The energy difference between the bright states and dark states $\delta_{0}$ and the FSS $\delta_{1}$ and $\delta_{2}$ of the bright and dark states are collectively referred as the exchange parameters, which are the vital parameters of X$^0$. To observe the dark states, the magnetic field in Voigt geometry is applied, which can couple the bright and the dark states and make the dark states optical accessible by destroying the rotational symmetry. We repeat the polarization-resolved PC measurements with the magnetic field from 0 to 4 T in Voigt geometry, as shown in Fig.~\ref{fig:4}(c). As expected, the dark states emerge at 1.5 T \cite{PhysRevApplied.8.064018}. The linear-polarization-resolved PC spectra at 4 T are shown in Fig.~\ref{fig:4}(d) with the blue and red short lines representing the dark and bright states respectively, which reveal the orthogonally linear polarizations clearly. The extracted energies of bright and dark states are shown in Fig.~\ref{fig:4}(b) by deducting the diamagnetic effect. But the fitting involves too many parameters, including the exchange parameters and in-plane g-factors of electron and hole. For more accuracy, we fit it with combining the results under Faraday magnetic field and a vector magnetic field, as will be discussed below.

\begin{figure}
\includegraphics[scale=0.5]{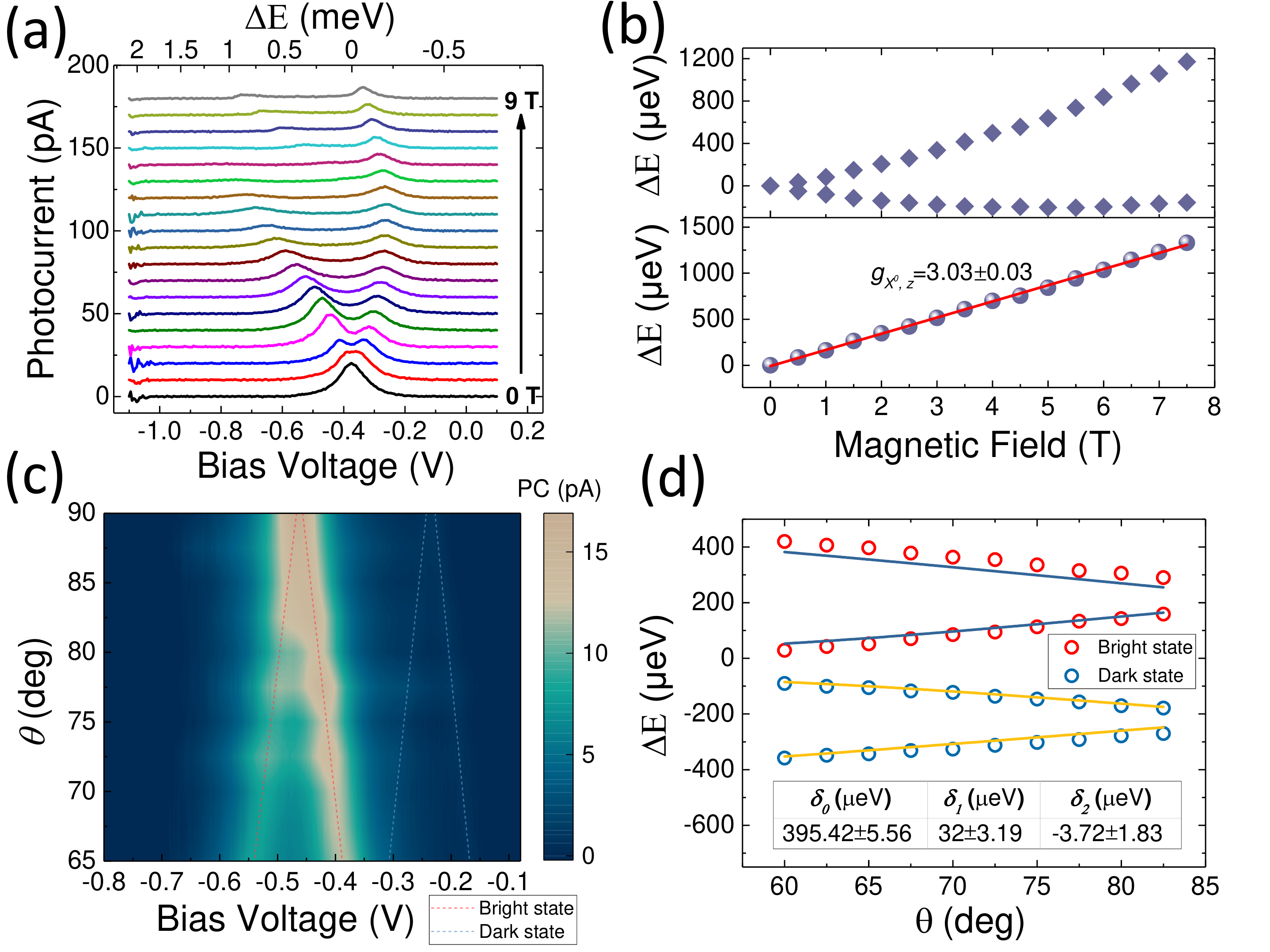}
\caption{\label{fig:5} (a) PC spectra of X$^0$ with a magnetic field from 0 to 9 T in Faraday geometry. The top x-axis represents the energy shift compared with the X$^+$ transition energy at zero magnetic field.  The spectra are shifted for clarity. (b) Top panel: The energy of X$^0$ extracted from (a) as a function of the magnetic field, offset by their mean at 0 T. Bottom panel: The Zeeman splitting of X$^0$, in which the g-factor is derived as $3.03\pm0.03$. (c) Magneto-PC mapping of X$^0$ under a 4 T vector magnetic field. The red and blue dashed lines represent the bright states and the dark states, respectively. (d) The fitting results of the splitting of X$^0$ under a vector magnetic field. The circles represent the extracted data of four transitions in (c) without diamagnetic shift. The solid curves show the fitted results. The fitted exchange parameters are shown in the inset table.}
\end{figure}

The measurements of the PC spectra of the X$^0$ under Faraday geometry are performed, which show Zeeman splitting and diamagnetic effect clearly, as presented in Fig.~\ref{fig:5}(a). The g-factor of X$^0$ is fitted as $3.03\pm0.03$, as shown in Fig.~\ref{fig:5}(b), which can be used as a more reliable parameter ($g_{X^0,z}=\vert{g_{e,z} + g_{h,z}}\vert$) to fit other parameters. Next, the PC spectra with a vector magnetic field at 4 T is performed, as shown in Fig.~\ref{fig:5}(c). For a small $\theta$ ranging from $75^{\circ}$ to $60^{\circ}$, which corresponds to a large magnetic field in Faraday geometry, the splittings of bright states and dark states increase due to the larger g-factor in Faraday geometry. On the other side, the PC spectra of dark excitons get weaker because of the weaker coupling between dark and bright excitons at a small magnetic field in Voigt geometry. We combine all the PC data with the magnetic field under different geometries and fit them together with the model in Ref.\onlinecite{PhysRevB.65.195315}, as shown in Fig.~\ref{fig:4}(b) and Fig.~\ref{fig:5}(d). The exchange parameters and g-factors are shown in Fig.~\ref{fig:5}(d) and Tables~\ref{tbl:example}, respectively. Due to the different Coulomb interactions, the g-factors have different values between X$^0$ and X$^+$. Also compared with X$^+$, the anisotropy factors for both electron and hole of X$^0$ are larger, which own to the different heavy hole-light hole mixing contribution in the different composition of the electronic states \cite{PhysRevB.77.241307}.

\begin{table}[h]
\small
  \caption{\ Values of g-factors and anisotropy factors for both positively charged X$^+$ and neutral exciton X$^0$ }
  \label{tbl:example}
  \centering
  \begin{tabular*}{0.48\textwidth}{@{\extracolsep{\fill}}lll}
    \hline
    \hline
    g-factor &  \multicolumn{1}{c}{X$^+$} & \multicolumn{1}{c}{X$^0$} \\
    \hline
    \multicolumn{1}{c}{$\vert{g_{e,x}}\vert$} & \multicolumn{1}{c}{$0.34\pm{0.01}$} & \multicolumn{1}{c}{$0.15\pm{0.03}$}  \\
    \multicolumn{1}{c}{$\vert{g_{h,x}}\vert$} & \multicolumn{1}{c}{$0.17\pm{0.01}$} & \multicolumn{1}{c}{$0.59\pm{0.10}$}  \\
    \multicolumn{1}{c}{$\vert{g_{e,z}}\vert$} & \multicolumn{1}{c}{$0.22\pm{0.05}$} & \multicolumn{1}{c}{$0.26\pm{0.08}$}  \\
    \multicolumn{1}{c}{$\vert{g_{h,z}}\vert$} & \multicolumn{1}{c}{$2.88\pm{0.02}$} & \multicolumn{1}{c}{$2.83\pm{0.15}$}  \\
    \multicolumn{1}{c}{$P_{e, x-z}$} & \multicolumn{1}{c}{$(17.91\pm{0.11})$\%} & \multicolumn{1}{c}{$(26.82\pm{0.34})$\%}  \\
    \multicolumn{1}{c}{$P_{h, x-z}$} & \multicolumn{1}{c}{$(88.55\pm{0.02})$\%} & \multicolumn{1}{c}{$(90\pm{0.17})$\%}  \\
    \hline
    \hline
  \end{tabular*}
\end{table}

It is worth noting that the fitting results of X$^0$ are not as good as those of X$^+$. There are many reasons. Firstly, the electron-hole exchange interaction and the existence of the FSS make the fitting of X$^0$ more complicated. Secondly, the linewidth of X$^0$ PC spectrum is much broader than that of X$^+$, which also increases the uncertainty of the measurement and the fitting. For example, the splitting under Voigt magnetic field cannot be distinguished directly. As a result, the polarization-resolved PC spectra measurements have to be performed and bring more errors. Finally, the PC amplitude of dark states is very weak and the splitting is small. So the related fitting parameters have relatively larger errors. Nevertheless, the PC spectroscopy of X$^0$ still provide a method to identify the exchange parameters and g-factor tensors of electron and hole.

\section{\label{sec:level1}Conclusion}

In conclusion, we measured the electron and hole g tensors of X$^+$ and X$^0$ in a single QD by the high resolution PC spectroscopy under a vector magnetic field. For X$^+$, the vertical and in-plane g-factors of electron and hole are obtained precisely. The PC spectrum of X$^+$ by two-color excitation also illustrates the opposite signs of electron and hole g-factors in Faraday geometry, showing the PC spectroscopy provides a convenient way to investigate spin properties. Furthermore, the large PC enhancement is observed with vector magnetic field when electron and hole have the same absolute value of g factor.
The g-factors of X$^0$, as well as the parameters related with electron-hole exchange interaction, are also fitted through the magneto-PC measurements. For both X$^+$ and X$^0$, the hole g-factors show larger anisotropy than those of electron. Our results prove that the PC spectroscopy is a powerful tool to study the QDs, and give the intuitive picture of the electron and hole g-factors of different excitons, which gives the full determination of the electronic properties of different excitonic states in QDs through the PC spectroscopy.

\begin{acknowledgments}

This work was supported by the National Natural Science Foundation of China under Grants No. 11934019, No. 61675228, No. 11721404, No. 51761145104 and No. 11874419; the Strategic Priority Research Program, the Instrument Developing Project and the Interdisciplinary Innovation Team of the Chinese Academy of Sciences under Grants No. XDB28000000 and No.YJKYYQ20180036, and the Key RD Program of Guangdong Province (Grant No.2018B030329001).

\end{acknowledgments}

\end{document}